\documentclass[aps,prl,twocolumn,superscriptaddress,showpacs]{revtex4-1}

\usepackage{graphicx}
\usepackage{dcolumn}
\usepackage{bm}

\begin{document}


\title{Experimental observation of the bifurcation dynamics of an intrinsic localized mode in a driven 1-D nonlinear lattice
}


\author{M. Sato}
\author{S. Imai}
\author{N. Fujita}
\author{S. Nishimura}
\author{Y. Takao}
\author{Y. Sada}
\affiliation{Graduate School of Natural Science and Technology, Kanazawa University,
Ishikawa 920-1192, Japan
}

\author{B. E. Hubbard}
\affiliation{Laboratory of Atomic and Solid State Physics, Cornell University,
Ithaca, NY 14853-2501, USA
}

\author{B. Ilic}

\affiliation{Cornell Nanoscale Science and Technology Facility, Cornell University,
Ithaca, NY 14853-2501, USA
}

\author{A. J. Sievers}
\affiliation{Laboratory of Atomic and Solid State Physics, Cornell University,
Ithaca, NY 14853-2501, USA
}

\date{\today}

\begin{abstract}
Linear response spectra of a driven intrinsic localized mode in a micromechanical array are measured as it approaches two fundamentally different kinds of bifurcation points. A linear phase mode associated with this autoresonant state softens in frequency and its amplitude grows as the upper frequency bifurcation point is approached, similar to the soft mode kinetic transition for a single driven Duffing resonator. A lower frequency bifurcation point occurs when the four-wave-mixing partner of this same phase mode intercepts the top of the extended wave branch, initiating a second kinetic transition process.
\end{abstract}


\maketitle

A general property of a driven nonlinear oscillator is that given sufficient starting amplitude it will stay in resonance as the frequency is changed adiabatically.  In this autoresonant (AR) state, where feedback is not required, the oscillator phase is locked to the driver. A variety of applications have been described in Ref.\cite{1,2}. Recent theoretical work has focused on controlling solitons\cite{3} and excitations in discrete and continuous nonlinear Schr\"{o}dinger equations\cite{4} while experimental studies have occurred for micromechanical arrays\cite{5}, optical guided waves\cite{6} and superconducting Josephson resonators\cite{7}. It has been predicted theoretically for a periodically driven single nonlinear oscillator and examined with analog electrical experiments near the location of its bifurcation region that critical phenomena arise in the density of fluctuations\cite{8,9}, and the bifurcation transition is characterized by slow dynamics\cite{10,11} associated with a soft mode\cite{12}, which goes to zero frequency at the transition.

Micro and nano-electrical-mechanical systems (MEMS and NEMS) provide a platform with which to study the intrinsic dynamical localization of vibrations in driven nonlinear lattices. Both experimental\cite{5,13} and theoretical\cite{5,14,15,16,17,18,19} studies have appeared that showcase the properties of such intrinsic localized modes (ILMs), the fundamental strongly localized excitation that appears when both lattice discreteness and nonlinearity are important\cite{20,21,22,23}. A feature overlooked until recently is the bifurcation properties of this autoresonant (AR) state and the concomitant back reaction of the driven ILM on the dynamics of the lattice. For a 1-D monatomic lattice with hard quartic anharmonicity it has been predicted that linear local modes (LLMs) would appear nearby an ILM and that four-wave mixing between the ILM and the LLM would give rise to additional spectral features\cite{24}. Missing are any experiments or discussion of the existence of a linear ILM soft phase mode or the influence of LLMs as the driven ILM approaches a bifurcation frequency.

In this work the first experimental observations of the dynamical properties of two separate bifurcation transitions from the AR-ILM state of a driven 1-D micromechanical array are presented and analyzed.  Linear response measurements are used to probe the small signal dynamics of AR state. Our measurements and associated simulations show that the high frequency bifurcation transition is very similar to that found for a single driven Duffing-like oscillator in that the associated linear soft phase mode goes to zero frequency at this transition\cite{25}. The low frequency transition, on the other hand, is characterized by a nonlinear interaction between the same soft mode and the highest frequency plane wave mode of the lattice.

\begin{figure}
\includegraphics{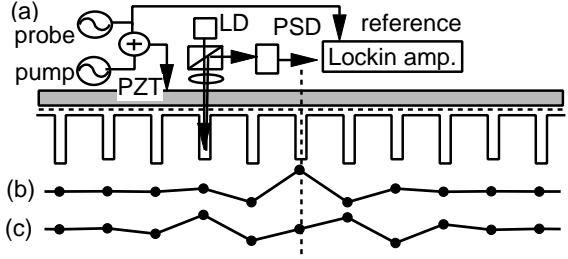}
\caption{\label{fig:fig1}(a) Experimental set up for the linear response measurement of the autoresonant (AR) ILM state with a uniform probe perturbation. The array is composed of alternating 50 and 55 $\mu $m length by 300 nm thick cantilevers. The pump driver and probe signals are added and used to excite the array uniformly with a PZT. The ILM position is monitored by a combination of a line focused laser beam and a 1-D CCD camera (not shown). A laser diode (LD) illuminates a short cantilever to the side of the ILM and the reflected beam is detected by a position sensitive detector (PSD). The displacement signal is analyzed with a lock-in amplifier. A typical pump amplitude is 14V, while the probe amplitude is 12 mV. (b) Spatial pattern of the odd symmetry ILM. (c) Pattern of an even linear local mode (LLM). The asymmetric LD heating of the localized vibration region permits the even-LLM to be excited even though the PZT drives the array uniformly.}
\end{figure}

Figure~\ref{fig:fig1} shows the experimental setup both for the AR amplitude measurement and for the associated linear response measurement. The driven micromechanical array contains 152 Si$_3$N$_4$ cantilevers coupled together by a common overhang. A cw pump feeds energy to the array maintaining the ILM in the large amplitude AR state. For linear response measurements an additional weak probe driver is used to perturb the array. The output of the probe is combined with the strong pump and connected to the PZT so the perturbation is applied uniformly across the lattice. First, the driver frequency is chirped up to generate the ILM at an arbitrary lattice site.  Its position is monitored by a combination of a line focused laser beam and a 1-D CCD camera (not shown). The probe laser beam is then adjusted to the next short cantilever of the ILM to operate in the small signal regime. The motion of this cantilever is monitored using a position sensitive detector(PSD) with a large deflection range. A lock-in amplifier is used to selectively analyze the cantilever motion that is caused by the probe oscillating at a given frequency. The response spectrum is measured by scanning the probe frequency, while the pump frequency is held fixed. By changing the pump frequency in a stepwise fashion, the linear mode properties can be monitored as a kinetic transition is approached. Once the driver frequency passes one of the bifurcation points, it takes many tries until an ILM is generated at the same lattice site for additional measurements with different pump frequencies. This uniform drive method should only couple to odd symmetry modes; however, heating from the probe laser produces an asymmetric perturbation near the ILM so that an even LLM of the type shown in Fig.~\ref{fig:fig1}(c) can also be observed.  

\begin{figure}
\includegraphics{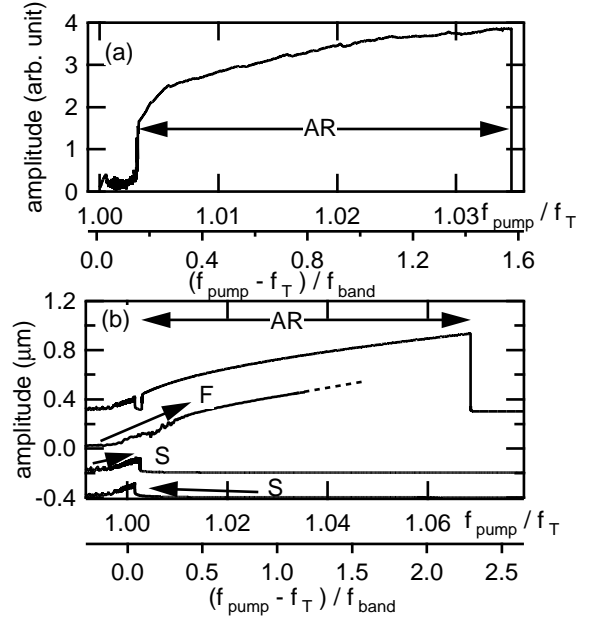}
\caption{\label{fig:fig2}(a) Experimentally observed AR ILM amplitude as a function of the pump frequency, showing two transitions. Upper abscissa: pump frequency normalized to the top of the optic branch. Lower abscissa: difference frequency between the pump and the top of the optic branch ($f_T=$140.0 kHz) normalized by the band width (3.1 kHz). The stable AR region is 140.46 kHz to 144.85 kHz, or 0.148 to 1.57 by the normalized difference frequency. (b) Simulated nonlinear response of the AR state in the hard nonlinear lattice with driver appropriate to the experimental level. Top trace: AR region, identifying its two transitions. Middle trace: F= fast frequency rate required to reach the AR state. Bottom two traces: S= slow up and down scanning, no AR state occurs. The stable frequency region is 137.56 to 146.54 kHz, or 0.102 to 2.29 by the normalized difference frequency. The top of the band frequency is 137.14 kHz and the band width is 4.1kHz. Curves are shifted up or down for clarity.}
\end{figure}

Both a high frequency and a low frequency bifurcation transition  out of the AR-ILM state are experimentally identified in Fig.~\ref{fig:fig2}(a) where the observed amplitude is shown as a function of the pump frequency. Because of the irreversible nature of the AR state beyond the transition points, a measurement sequence is initiated from its middle frequency region and the pump frequency is then step incremented slowly up or down. The top abscissa in Fig.~\ref{fig:fig2}(a) is the driving frequency normalized to the top of the linear optic mode frequency. The lower abscissa is the difference frequency between the driver and the linear optic mode frequency  normalized by the optical band width. (This ratio provides a general measure of the strength of the nonlinearity for this driven system.) Note that to reach the high amplitude AR state the pump frequency must increase at a sufficient rate to cross over the low amplitude state between the unlocked state and the AR state\cite{17}. 

These experimental results are compared to simulations incorporating previously applied lumped element lattice model equations
\[
\begin{array}{l}
 m_i \ddot x_i  + m_i \dot x_i /\tau  + k_{2Oi} x_i  + \sum\limits_j {k_{2I}^{(j)} } \left( {2x_i  - x_{i + j}  - x_{i - j} } \right) \\ 
  + k_{4O} x_i^3  + k_{4I} \left\{ {\left( {x_i  - x_{i + 1} } \right)^3  + \left( {x_i  - x_{i - 1} } \right)^3 } \right\} \\ 
  = m_i \alpha _{pump} \cos \Omega t + m_i \alpha _{probe} \cos \omega t \\ 
 \end{array}
\]
with an obvious notation. Specific array parameters are described in the Supplemental Material\cite{supp}. The top curve in Fig.~\ref{fig:fig2}(b) is for a driver with amplitude similar to the experimental level. Once the AR-ILM state is generated, its amplitude changes smoothly with a slow variation in the driver frequency. Both AR transitions are evident although the lower frequency one is not as marked as observed experimentally in Fig.~\ref{fig:fig2}(a). The next lower trace in Fig.~\ref{fig:fig2}(b) shows the result of a relatively rapid increase in the driver frequency necessary to produce the AR state. The bottom two traces illustrate the small amplitudes that appear for slow up and down frequency scanning where no AR state is produced.

\begin{figure}
\includegraphics[width=7cm]{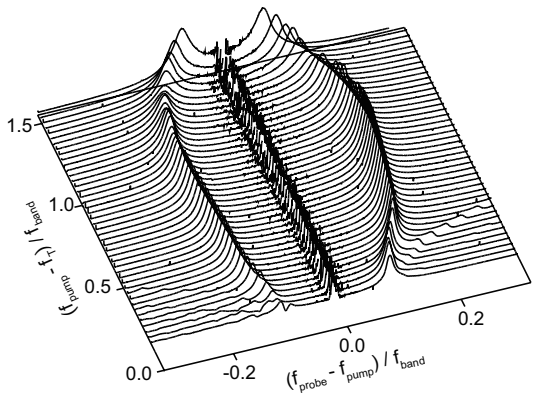}
\caption{\label{fig:fig3}Experimental response spectra for the AR-ILM state produced by a uniform driver. The two strong sidebands identify the soft phase mode associated with this state. Spectra are aligned from (0.161 to 1.55 by the normalized difference frequency). The upper and lower frequency limits are near the two bifurcation frequencies. The gap frequency of these sidebands decreases and the response grows as the pump frequency approaches the upper bifurcation point. The additional very weak sidebands approaching the lower bifurcation transition are the even LLM and its partner and lower frequency band modes, as described in the text .Both frequency axes are normalized by the optical band width.}
\end{figure}

The experimentally measured linear response spectra for the AR-ILM state at different pump frequencies are presented in Fig.~\ref{fig:fig3}. As will be demonstrated via simulations below, the two strong sidebands shown here are due to the soft phase mode within the AR state. (For clarity the actual pumped ILM amplitude is deleted in this figure.) The probe spectra are displayed with the pump varying from 140.5 kHz to 144.8 kHz in 100 Hz intervals from bottom to top. (This range corresponds to 0.161-1.55 in terms of the difference frequency normalized by the band width.) The higher frequency sideband is the soft phase mode while the lower frequency one is its four wave mixing partner. The two pump frequency limits shown are close to the upper and lower bifurcation frequencies, and so the frequency range essentially corresponds to the entire stable region of the AR state shown in Fig.~\ref{fig:fig2}(a). Note the characteristic soft mode behavior as the pump frequency approaches the upper bifurcation point. The upper bifurcation transition takes place when the soft phase mode frequency goes to zero. Similar soft phase mode behavior is found for a single driven Duffing oscillator\cite{25}. Most of the weak satellite features that appear near the low frequency bifurcation transition are extended wave optic modes and their partners but the highest frequency one is an LLM.

\begin{figure}
\includegraphics[width=7cm]{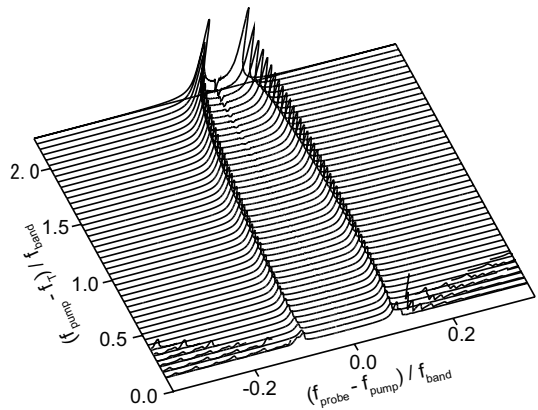}
\caption{\label{fig:fig4}Simulated linear response spectra of the AR state produced by a uniform driver. The entire stable frequency region of the pump frequency is shown, i.e., 0.112 to 2.26 by the normalized difference frequency. At the upper bifurcation point, the phase mode gap frequency softens and response diverges. At the lower bifurcation point, the phase mode overlaps with the top band mode, shown as small peaks near the bottom of the figure. This AR transition occurs when the soft phase mode intersects these linear modes via 4-wave mixing. The band mode maxima near the lower bifurcation region are magnified 20 times. }
\end{figure}

From numerical simulations the corresponding linear response spectra for the same uniform perturbation applied to the AR-ILM state can be obtained.The displacement produced by the perturbation is recorded in a manner similar to that of experiment, i.e., the displacement is multiplied by $\cos \omega t$ or $\sin \omega t$ and averaged over time to obtain cosine and sine components of the response function. The resulting amplitude spectra calculated from these cosine and sine spectra are shown in Fig.~\ref{fig:fig4}. The entire stable pump frequency region (0.112-2.26) for the AR-ILM state is shown. As the upper bifurcation point is approached, the phase mode frequency softens and its response diverges. The associated LLM (off the horizontal scale) has no such dramatic behavior. As the lower bifurcation point is approached the four wave mixing partner of the soft phase mode approaches the top of this extended wave branch. The optic branch modes and their four wave mixing partners appear as weak low frequency satellite features in the figure. The lower transition occurs when these two intersect, presenting a completely different dynamical signature from that of the upper transition. The detailed properties of the modes in Fig.~\ref{fig:fig4} are identified by checking their shapes numerically. We find that the odd symmetry of the soft phase mode is the same as that of the driven ILM. The other small peaks in Fig.~\ref{fig:fig4} are odd symmetry band modes.

\begin{figure}
\includegraphics{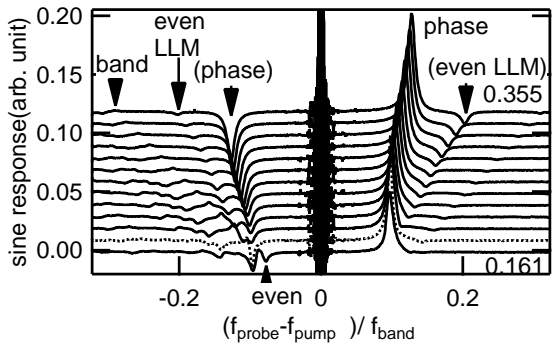}
\caption{\label{fig:fig5}Sine component of the experimental probe response near by the lower bifurcation frequency.  Spectra are ordered by the pump frequency from 0.355 to 0.161 in 0.016 step. The large peak at the center is the pump signal. The lower bifurcation takes place at 0.148. The large positive amplitude identifies the soft phase mode oscillation of the ILM, the corresponding negative peak is its four wave mixing partner. The even LLM, activated by asymmetric heating of the ILM, appears as an amplitude modulation response. As the driver frequency is decreased the highest frequency band mode of the array, identified with an arrow, approaches the partner of the soft phase mode. (Lower frequency band modes appear as a sequence of positive and negative peaks.) Near the transition the even LLM frequency crosses the soft phase mode partner, dashed curve, second from the bottom. The transition takes place when the soft mode partner intersects the lower frequency odd symmetry band mode.}
\end{figure}

The dynamics behind the lower bifurcation transition is quite different from the upper one. The experimentally measured response for these different excitations is shown in Fig.~\ref{fig:fig5}. The sine component of the probe response spectra is displayed as the lower bifurcation transition is approached. The large positive and negative peaks are signatures of  the soft phase mode and its partner. The top of the plane wave vibrational band is evident because it is followed by many smaller positive and negative peaks at still larger frequency shifts. The even symmetry LLM mode is activated by heating from the focused laser probe beam to one side of the ILM and gives an AM modulation component. Our experiments show that as the lower bifurcation transition is approached this LLM mode crosses over the partner of the soft phase mode, as shown in Fig.~\ref{fig:fig5}, with no effect. Only when this soft mode partner coalesces with the top of the odd symmetry band mode spectrum does the transition occur.

Our experimental measurements of  the linear response spectrum of the AR-ILM state in a lattice show that a soft phase mode and its four wave mixing partner play a key role in the bifurcation transitions. The soft mode frequency collapses to zero and its amplitude diverges at the upper bifurcation point, signifying a kinetic phase transition very similar to that observed for a single driven Duffing oscillator. The LLMs attached to the AR state show no such characteristic frequency behavior. With decreasing frequency and amplitude of this AR-ILM state some LLMs transform to delocalized optic modes while one does not before the new, lower frequency, bifurcation transition occurs. As this transition is approached it is the resonant four wave mixing of the soft phase mode with the delocalized optic mode, of the same symmetry, that destroys the localized AR-ILM state by modulating its shape with a breathing oscillation of increasing amplitude. Although these experiments are specific to a micromechanical array these  results are expected to be quite general with the multiple bifurcation nonlinear dynamics applicable to other kinds of physical lattices. 

\begin{acknowledgments}
The research was supported by Grant-in-Aid for challenging Exploratory Research (21656022). BEH and AJS were supported by NSF-DMR-0906491 and DOE-DE-FG02-04ER46154.
\end{acknowledgments}


\begin{thebibliography}{26}%
\makeatletter
\providecommand \@ifxundefined [1]{%
 \@ifx{#1\undefined}
}%
\providecommand \@ifnum [1]{%
 \ifnum #1\expandafter \@firstoftwo
 \else \expandafter \@secondoftwo
 \fi
}%
\providecommand \@ifx [1]{%
 \ifx #1\expandafter \@firstoftwo
 \else \expandafter \@secondoftwo
 \fi
}%
\providecommand \natexlab [1]{#1}%
\providecommand \enquote  [1]{``#1''}%
\providecommand \bibnamefont  [1]{#1}%
\providecommand \bibfnamefont [1]{#1}%
\providecommand \citenamefont [1]{#1}%
\providecommand \href@noop [0]{\@secondoftwo}%
\providecommand \href [0]{\begingroup \@sanitize@url \@href}%
\providecommand \@href[1]{\@@startlink{#1}\@@href}%
\providecommand \@@href[1]{\endgroup#1\@@endlink}%
\providecommand \@sanitize@url [0]{\catcode `\\12\catcode `\$12\catcode
  `\&12\catcode `\#12\catcode `\^12\catcode `\_12\catcode `\%12\relax}%
\providecommand \@@startlink[1]{}%
\providecommand \@@endlink[0]{}%
\providecommand \url  [0]{\begingroup\@sanitize@url \@url }%
\providecommand \@url [1]{\endgroup\@href {#1}{\urlprefix }}%
\providecommand \urlprefix  [0]{URL }%
\providecommand \Eprint [0]{\href }%
\providecommand \doibase [0]{http://dx.doi.org/}%
\providecommand \selectlanguage [0]{\@gobble}%
\providecommand \bibinfo  [0]{\@secondoftwo}%
\providecommand \bibfield  [0]{\@secondoftwo}%
\providecommand \translation [1]{[#1]}%
\providecommand \BibitemOpen [0]{}%
\providecommand \bibitemStop [0]{}%
\providecommand \bibitemNoStop [0]{.\EOS\space}%
\providecommand \EOS [0]{\spacefactor3000\relax}%
\providecommand \BibitemShut  [1]{\csname bibitem#1\endcsname}%
\let\auto@bib@innerbib\@empty
\bibitem [{\citenamefont {Fajans}\ \emph {et~al.}(1999)\citenamefont {Fajans},
  \citenamefont {Gilson},\ and\ \citenamefont {Friedland}}]{1}%
  \BibitemOpen
  \bibfield  {author} {\bibinfo {author} {\bibfnamefont {J.}~\bibnamefont
  {Fajans}}, \bibinfo {author} {\bibfnamefont {E.}~\bibnamefont {Gilson}}, \
  and\ \bibinfo {author} {\bibfnamefont {L.}~\bibnamefont {Friedland}},\
  }\href@noop {} {\bibfield  {journal} {\bibinfo  {journal} {Phys. Plasmas}\
  }\textbf {\bibinfo {volume} {6}},\ \bibinfo {pages} {4497} (\bibinfo {year}
  {1999})}\BibitemShut {NoStop}%
\bibitem [{\citenamefont {Fajans}\ and\ \citenamefont {Friedland}(2001)}]{2}%
  \BibitemOpen
  \bibfield  {author} {\bibinfo {author} {\bibfnamefont {J.}~\bibnamefont
  {Fajans}}\ and\ \bibinfo {author} {\bibfnamefont {L.}~\bibnamefont
  {Friedland}},\ }\href@noop {} {\bibfield  {journal} {\bibinfo  {journal} {Am.
  J. Phys.}\ }\textbf {\bibinfo {volume} {69}},\ \bibinfo {pages} {1096}
  (\bibinfo {year} {2001})}\BibitemShut {NoStop}%
\bibitem [{\citenamefont {Batalov}\ and\ \citenamefont {Shagalov}(2011)}]{3}%
  \BibitemOpen
  \bibfield  {author} {\bibinfo {author} {\bibfnamefont {S.~V.}\ \bibnamefont
  {Batalov}}\ and\ \bibinfo {author} {\bibfnamefont {A.~G.}\ \bibnamefont
  {Shagalov}},\ }\href@noop {} {\bibfield  {journal} {\bibinfo  {journal}
  {Phys. Rev. E}\ }\textbf {\bibinfo {volume} {84}},\ \bibinfo {pages} {016603}
  (\bibinfo {year} {2011})}\BibitemShut {NoStop}%
\bibitem [{\citenamefont {Gopher}\ \emph {et~al.}(2005)\citenamefont {Gopher},
  \citenamefont {Friedlander},\ and\ \citenamefont {Shagalov}}]{4}%
  \BibitemOpen
  \bibfield  {author} {\bibinfo {author} {\bibfnamefont {Y.}~\bibnamefont
  {Gopher}}, \bibinfo {author} {\bibfnamefont {L.}~\bibnamefont {Friedlander}},
  \ and\ \bibinfo {author} {\bibfnamefont {A.~G.}\ \bibnamefont {Shagalov}},\
  }\href@noop {} {\bibfield  {journal} {\bibinfo  {journal} {Phys. Rev. E}\
  }\textbf {\bibinfo {volume} {72}},\ \bibinfo {pages} {036604} (\bibinfo
  {year} {2005})}\BibitemShut {NoStop}%
\bibitem [{\citenamefont {Sato}\ \emph {et~al.}(2006)\citenamefont {Sato},
  \citenamefont {Hubbard},\ and\ \citenamefont {Sievers}}]{5}%
  \BibitemOpen
  \bibfield  {author} {\bibinfo {author} {\bibfnamefont {M.}~\bibnamefont
  {Sato}}, \bibinfo {author} {\bibfnamefont {B.~E.}\ \bibnamefont {Hubbard}}, \
  and\ \bibinfo {author} {\bibfnamefont {A.~J.}\ \bibnamefont {Sievers}},\
  }\href@noop {} {\bibfield  {journal} {\bibinfo  {journal} {Rev. Mod. Phys.}\
  }\textbf {\bibinfo {volume} {78}},\ \bibinfo {pages} {137} (\bibinfo {year}
  {2006})}\BibitemShut {NoStop}%
\bibitem [{\citenamefont {{A. Barak {\it {et~al.}}}}(2009)}]{6}%
  \BibitemOpen
  \bibfield  {author} {\bibinfo {author} {\bibnamefont {{A. Barak {\it
  {et~al.}}}}},\ }\href@noop {} {\bibfield  {journal} {\bibinfo  {journal}
  {Phys. Rev. Lett.}\ }\textbf {\bibinfo {volume} {103}},\ \bibinfo {pages}
  {123901} (\bibinfo {year} {2009})}\BibitemShut {NoStop}%
\bibitem [{\citenamefont {{O. Naaman {\it {et~al.}}}}(2008)}]{7}%
  \BibitemOpen
  \bibfield  {author} {\bibinfo {author} {\bibnamefont {{O. Naaman {\it
  {et~al.}}}}},\ }\href@noop {} {\bibfield  {journal} {\bibinfo  {journal}
  {Phys. Rev. Lett.}\ }\textbf {\bibinfo {volume} {101}},\ \bibinfo {pages}
  {117005} (\bibinfo {year} {2008})}\BibitemShut {NoStop}%
\bibitem [{\citenamefont {{M. I. Dykman {\it {et~al.}}}}(1994)}]{8}%
  \BibitemOpen
  \bibfield  {author} {\bibinfo {author} {\bibnamefont {{M. I. Dykman {\it
  {et~al.}}}}},\ }\href@noop {} {\bibfield  {journal} {\bibinfo  {journal}
  {Phys. Rev. E}\ }\textbf {\bibinfo {volume} {49}},\ \bibinfo {pages} {1198}
  (\bibinfo {year} {1994})}\BibitemShut {NoStop}%
\bibitem [{\citenamefont {Luchinsky}\ \emph {et~al.}(1998)\citenamefont
  {Luchinsky}, \citenamefont {McClintock},\ and\ \citenamefont {Dykman}}]{9}%
  \BibitemOpen
  \bibfield  {author} {\bibinfo {author} {\bibfnamefont {D.~G.}\ \bibnamefont
  {Luchinsky}}, \bibinfo {author} {\bibfnamefont {P.~V.~E.}\ \bibnamefont
  {McClintock}}, \ and\ \bibinfo {author} {\bibfnamefont {M.~I.}\ \bibnamefont
  {Dykman}},\ }\href@noop {} {\bibfield  {journal} {\bibinfo  {journal} {Rep.
  Prog. Phys.}\ }\textbf {\bibinfo {volume} {61}},\ \bibinfo {pages} {889}
  (\bibinfo {year} {1998})}\BibitemShut {NoStop}%
\bibitem [{\citenamefont {Guckenheimer}\ and\ \citenamefont
  {P.Holmes}(1983)}]{10}%
  \BibitemOpen
  \bibfield  {author} {\bibinfo {author} {\bibfnamefont {J.}~\bibnamefont
  {Guckenheimer}}\ and\ \bibinfo {author} {\bibnamefont {P.Holmes}},\
  }\href@noop {} {\emph {\bibinfo {title} {Nonlinear Oscillations, Dynamical
  Systems, and Bifurcations of Vector Fields}}}\ (\bibinfo  {publisher}
  {Springer-Verlag},\ \bibinfo {year} {1983})\BibitemShut {NoStop}%
\bibitem [{\citenamefont {{D. Ryvkine and M. I. Dykman, and B.
  Golding}}(2004)}]{11}%
  \BibitemOpen
  \bibfield  {author} {\bibinfo {author} {\bibnamefont {{D. Ryvkine and M. I.
  Dykman, and B. Golding}}},\ }\href@noop {} {\bibfield  {journal} {\bibinfo
  {journal} {Phys. Rev. E}\ }\textbf {\bibinfo {volume} {69}},\ \bibinfo
  {pages} {061102} (\bibinfo {year} {2004})}\BibitemShut {NoStop}%
\bibitem [{\citenamefont {Thompson}\ and\ \citenamefont {Virgin}(1986)}]{12}%
  \BibitemOpen
  \bibfield  {author} {\bibinfo {author} {\bibfnamefont {J.~M.~T.}\
  \bibnamefont {Thompson}}\ and\ \bibinfo {author} {\bibfnamefont {L.~N.}\
  \bibnamefont {Virgin}},\ }\href@noop {} {\bibfield  {journal} {\bibinfo
  {journal} {Int. J. Non-Linear Mech.}\ }\textbf {\bibinfo {volume} {21}},\
  \bibinfo {pages} {205} (\bibinfo {year} {1986})}\BibitemShut {NoStop}%
\bibitem [{\citenamefont {{M. Spletzer {\it {et~al.}}}}(2006)}]{13}%
  \BibitemOpen
  \bibfield  {author} {\bibinfo {author} {\bibnamefont {{M. Spletzer {\it
  {et~al.}}}}},\ }\href@noop {} {\bibfield  {journal} {\bibinfo  {journal}
  {Appl. Phys. Lett.}\ }\textbf {\bibinfo {volume} {88}},\ \bibinfo {pages}
  {254102} (\bibinfo {year} {2006})}\BibitemShut {NoStop}%
\bibitem [{\citenamefont {Maniadis}\ and\ \citenamefont {Flach}(2006)}]{14}%
  \BibitemOpen
  \bibfield  {author} {\bibinfo {author} {\bibfnamefont {P.}~\bibnamefont
  {Maniadis}}\ and\ \bibinfo {author} {\bibfnamefont {S.}~\bibnamefont
  {Flach}},\ }\href@noop {} {\bibfield  {journal} {\bibinfo  {journal}
  {Europhys. Lett.}\ }\textbf {\bibinfo {volume} {74}},\ \bibinfo {pages} {452}
  (\bibinfo {year} {2006})}\BibitemShut {NoStop}%
\bibitem [{\citenamefont {Dick}\ \emph {et~al.}(2008)\citenamefont {Dick},
  \citenamefont {Balachandran},\ and\ \citenamefont {Mote}}]{15}%
  \BibitemOpen
  \bibfield  {author} {\bibinfo {author} {\bibfnamefont {A.~J.}\ \bibnamefont
  {Dick}}, \bibinfo {author} {\bibfnamefont {A.~J.}\ \bibnamefont
  {Balachandran}}, \ and\ \bibinfo {author} {\bibfnamefont {C.~D.}\
  \bibnamefont {Mote}},\ }\href@noop {} {\bibfield  {journal} {\bibinfo
  {journal} {Nonlin. Dyn.}\ }\textbf {\bibinfo {volume} {54}},\ \bibinfo
  {pages} {13} (\bibinfo {year} {2008})}\BibitemShut {NoStop}%
\bibitem [{\citenamefont {Wiersig}\ \emph {et~al.}(2009)\citenamefont
  {Wiersig}, \citenamefont {Flach},\ and\ \citenamefont {Ahn}}]{16}%
  \BibitemOpen
  \bibfield  {author} {\bibinfo {author} {\bibfnamefont {J.}~\bibnamefont
  {Wiersig}}, \bibinfo {author} {\bibfnamefont {S.}~\bibnamefont {Flach}}, \
  and\ \bibinfo {author} {\bibfnamefont {K.~H.}\ \bibnamefont {Ahn}},\
  }\href@noop {} {\bibfield  {journal} {\bibinfo  {journal} {Appl. Phys.
  Lett.}\ }\textbf {\bibinfo {volume} {93}},\ \bibinfo {pages} {222110}
  (\bibinfo {year} {2009})}\BibitemShut {NoStop}%
\bibitem [{\citenamefont {{Q. Chen {\it {et~al.}}}}(2009)}]{17}%
  \BibitemOpen
  \bibfield  {author} {\bibinfo {author} {\bibnamefont {{Q. Chen {\it
  {et~al.}}}}},\ }\href@noop {} {\bibfield  {journal} {\bibinfo  {journal}
  {Chaos}\ }\textbf {\bibinfo {volume} {19}},\ \bibinfo {pages} {013127}
  (\bibinfo {year} {2009})}\BibitemShut {NoStop}%
\bibitem [{\citenamefont {Kenig}\ \emph {et~al.}(2009)\citenamefont {Kenig},
  \citenamefont {Lifshitz},\ and\ \citenamefont {Cross}}]{18}%
  \BibitemOpen
  \bibfield  {author} {\bibinfo {author} {\bibfnamefont {E.}~\bibnamefont
  {Kenig}}, \bibinfo {author} {\bibfnamefont {R.}~\bibnamefont {Lifshitz}}, \
  and\ \bibinfo {author} {\bibfnamefont {M.~C.}\ \bibnamefont {Cross}},\
  }\href@noop {} {\bibfield  {journal} {\bibinfo  {journal} {Phys. Rev. E}\
  }\textbf {\bibinfo {volume} {79}},\ \bibinfo {pages} {026203} (\bibinfo
  {year} {2009})}\BibitemShut {NoStop}%
\bibitem [{\citenamefont {{E. Kenig {\it {et~al.}}}}(2009)}]{19}%
  \BibitemOpen
  \bibfield  {author} {\bibinfo {author} {\bibnamefont {{E. Kenig {\it
  {et~al.}}}}},\ }\href@noop {} {\bibfield  {journal} {\bibinfo  {journal}
  {Phys. Rev. E}\ }\textbf {\bibinfo {volume} {80}},\ \bibinfo {pages} {046202}
  (\bibinfo {year} {2009})}\BibitemShut {NoStop}%
\bibitem [{\citenamefont {Kiselev}\ \emph {et~al.}(1995)\citenamefont
  {Kiselev}, \citenamefont {Bickham},\ and\ \citenamefont {Sievers}}]{20}%
  \BibitemOpen
  \bibfield  {author} {\bibinfo {author} {\bibfnamefont {S.~A.}\ \bibnamefont
  {Kiselev}}, \bibinfo {author} {\bibfnamefont {S.~R.}\ \bibnamefont
  {Bickham}}, \ and\ \bibinfo {author} {\bibfnamefont {A.~J.}\ \bibnamefont
  {Sievers}},\ }\href@noop {} {\bibfield  {journal} {\bibinfo  {journal} {Comm.
  in Cond. Mat. Phys.}\ }\textbf {\bibinfo {volume} {17}},\ \bibinfo {pages}
  {135} (\bibinfo {year} {1995})}\BibitemShut {NoStop}%
\bibitem [{\citenamefont {Lai}\ and\ \citenamefont {Sievers}(1999)}]{21}%
  \BibitemOpen
  \bibfield  {author} {\bibinfo {author} {\bibfnamefont {R.}~\bibnamefont
  {Lai}}\ and\ \bibinfo {author} {\bibfnamefont {A.~J.}\ \bibnamefont
  {Sievers}},\ }\href@noop {} {\bibfield  {journal} {\bibinfo  {journal} {Phys.
  Repts.}\ }\textbf {\bibinfo {volume} {314}},\ \bibinfo {pages} {147}
  (\bibinfo {year} {1999})}\BibitemShut {NoStop}%
\bibitem [{\citenamefont {Campbell}\ \emph {et~al.}(2004)\citenamefont
  {Campbell}, \citenamefont {Flach},\ and\ \citenamefont {Kivshar}}]{22}%
  \BibitemOpen
  \bibfield  {author} {\bibinfo {author} {\bibfnamefont {D.~K.}\ \bibnamefont
  {Campbell}}, \bibinfo {author} {\bibfnamefont {S.}~\bibnamefont {Flach}}, \
  and\ \bibinfo {author} {\bibfnamefont {Y.~S.}\ \bibnamefont {Kivshar}},\
  }\href@noop {} {\bibfield  {journal} {\bibinfo  {journal} {Physics Today}\
  }\textbf {\bibinfo {volume} {57}},\ \bibinfo {pages} {43} (\bibinfo {year}
  {2004})}\BibitemShut {NoStop}%
\bibitem [{\citenamefont {Flach}\ and\ \citenamefont {Gorbach}(2008)}]{23}%
  \BibitemOpen
  \bibfield  {author} {\bibinfo {author} {\bibfnamefont {S.}~\bibnamefont
  {Flach}}\ and\ \bibinfo {author} {\bibfnamefont {A.}~\bibnamefont
  {Gorbach}},\ }\href@noop {} {\bibfield  {journal} {\bibinfo  {journal} {Phys.
  Repts.}\ }\textbf {\bibinfo {volume} {467}},\ \bibinfo {pages} {1} (\bibinfo
  {year} {2008})}\BibitemShut {NoStop}%
\bibitem [{\citenamefont {{V. Hizhnyakov {\it {et~al.}}}}(2006)}]{24}%
  \BibitemOpen
  \bibfield  {author} {\bibinfo {author} {\bibnamefont {{V. Hizhnyakov {\it
  {et~al.}}}}},\ }\href@noop {} {\bibfield  {journal} {\bibinfo  {journal}
  {Phys. Rev. B}\ }\textbf {\bibinfo {volume} {73}},\ \bibinfo {pages} {224302}
  (\bibinfo {year} {2006})}\BibitemShut {NoStop}%
\bibitem [{\citenamefont {Thompson}\ and\ \citenamefont {Stewart}(1987)}]{25}%
  \BibitemOpen
  \bibfield  {author} {\bibinfo {author} {\bibfnamefont {J.~M.~T.}\
  \bibnamefont {Thompson}}\ and\ \bibinfo {author} {\bibfnamefont {H.~B.}\
  \bibnamefont {Stewart}},\ }\href@noop {} {\emph {\bibinfo {title} {Nonlinear
  Dynamics and Chaos}}}\ (\bibinfo  {publisher} {John Wiley \& Sons},\ \bibinfo
  {year} {1987})\BibitemShut {NoStop}%
\bibitem [{sup()}]{supp}%
  \BibitemOpen
  \href@noop {} {}\bibinfo {note} {Parameters are in 2nd column of Table II in
  Ref.\cite{5}. See Supplemental Material at [URL will be inserted by
  publisher] for details.}\BibitemShut {Stop}%
\end{thebibliography}

\providecommand{\noopsort}[1]{}\providecommand{\singleletter}[1]{#1}%

\end{document}